\documentclass[aps,prx,reprint]{revtex4-1}

\usepackage[mathlines]{lineno}
\usepackage{lipsum}
\usepackage{blindtext}
\usepackage[export]{adjustbox}
\usepackage{graphicx,epsfig,epstopdf}
\usepackage{amsmath,amssymb}
\usepackage{esvect}
\usepackage{mathtools}
\usepackage{color}
\usepackage{subfigure}
\usepackage{array}
\usepackage{tabularx}
\usepackage{multirow}
\usepackage{booktabs}
\usepackage{mathtools}
\usepackage{float}
\newcolumntype{L}[1]{>{\raggedright\arraybackslash}p{#1}}
\newcolumntype{C}[1]{>{\centering\arraybackslash}p{#1}}
\newcolumntype{M}[1]{>{\centering\arraybackslash}m{#1}}
\allowdisplaybreaks

\def\_#1{{\bf #1}}

\def\.{\cdot}

\def\l#1{\label{eq:#1}}
\def\r#1{(\ref{eq:#1})}
\def\aee{\overline{\overline{\alpha}}_{\rm ee}}
\def\aem{\overline{\overline{\alpha}}_{\rm em}}
\def\ame{\overline{\overline{\alpha}}_{\rm me}}
\def\amm{\overline{\overline{\alpha}}_{\rm mm}}

\def\aeer{\overline{\overline{\alpha}}_{\rm ee, r}}
\def\aeen{\overline{\overline{\alpha}}_{\rm ee, n}}
\def\aemr{\overline{\overline{\alpha}}_{\rm em, r}}
\def\aemn{\overline{\overline{\alpha}}_{\rm em, n}}

\def\ammr{\overline{\overline{\alpha}}_{\rm mm, r}}
\def\ammn{\overline{\overline{\alpha}}_{\rm mm, n}}

\def\e{\begin{equation}}
\def\f{\end{equation}}

\def\eexx{\alpha^{\rm xx}_{\rm ee}}
\def\eexy{\alpha^{\rm xy}_{\rm ee}}
\def\eeyx{\alpha^{\rm yx}_{\rm ee}}
\def\eeyy{\alpha^{\rm yy}_{\rm ee}}
\def\eexz{\alpha^{\rm xz}_{\rm ee}}
\def\eezx{\alpha^{\rm zx}_{\rm ee}}
\def\eezz{\alpha^{\rm zz}_{\rm ee}}
\def\eeyz{\alpha^{\rm yz}_{\rm ee}}
\def\eezy{\alpha^{\rm zy}_{\rm ee}}

\def\H0{{H_0}}
\def\E0{\eta_0 {H_0}}

\def\=#1{\overline{\overline #1}}

\begin{document}

\title{Materiatronics: Modular analysis of arbitrary meta-atoms}

\author{Viktar S. Asadchy$^{1,2}$}
\author{Sergei A. Tretyakov$^{1}$}

\affiliation{$^1$Department of Electronics and Nanoengineering, 
Aalto University, P.~O.~Box~15500, FI-00076 Aalto, Finland\\
$^2$Department of General Physics, Francisk Skorina Gomel State University, 246019 Gomel, Belarus}

\begin{abstract}
Within the paradigm of metamaterials and metasurfaces, electromagnetic properties of composite materials can be engineered by shaping or modulating   their constituents, so-called meta-atoms. Synthesis and analysis of complex-shape meta-atoms with general polarization properties is a challenging task. 
In this paper, we demonstrate that the most general  response can be conceptually    decomposed into a set of basic,  fundamental polarization phenomena, which   enables immediate all-direction characterization of electromagnetic properties of arbitrary linear materials and metamaterials.  The proposed platform of modular characterization (called ``materiatronics'')  is  tested on several examples of bianisotropic and nonreciprocal  meta-atoms.  As a demonstration of the potential of the modular analysis,  we use it  to design a single-layer  metasurface of vanishing thickness with record-breaking unitary circular dichroism. 
The  analysis approach developed in this paper is supported by a ready-to-use computational code and can be further extended to meta-atoms engineered for other types of wave interactions, such as acoustics and mechanics.

\end{abstract}

\maketitle

\section{Introduction}
Metamaterials are engineered composites consisting of   tailored subwavelength meta-atoms (inclusions) for controlling wave phenomena at will. During last two decades, metamaterials have attracted great interest of researchers working in different fields of physics, which resulted in important impact on fundamental science and emergence of numerous fascinating concepts~\cite{smith_metamaterials_2004,liu_micromachined_2012,
cui_microwave_2018,kivshar_all-dielectric_2018,sklan_thermal_2018}. 
Nevertheless, there is no universal approach for synthesis of arbitrary complex metamaterials, mainly due to the two following reasons. First, determining the required  properties of single meta-atoms in the composite is a complicated inverse problem due to  their mutual interactions. Second, even if such properties are known, practical realization of meta-atoms can be still very challenging because it implies rigorous design and optimization of their anisotropic properties, i.e. their entire polarizability tensors. For this reason, the vast majority of works in the literature are  devoted to metamaterials whose functionality is  engineered only for a specific and small number (usually single)  of illumination directions.
Meanwhile,  properties of  metamaterials excited by fields of other configurations  remain  unknown and unprescribed. However, this knowledge is often very important and can lead to  unexpected and unique phenomena, as it was demonstrated by the example of  planar  chirality~\cite{bunn_chemical_1961,williams_opticalrotatory_1969,
sochava_chiral_1997,papakostas_optical_2003,
plum_metamaterials:_2009} which occurs only at specific oblique-angle illuminations in contrast to true three-dimensional chirality. 
Furthermore, in many situations, one needs to know under what excitation  a meta-atom will  respond in some particular way, for example, exhibiting highest possible Willis coupling~\cite{quan_maximum_2018}, strongest mechanical twist~\cite{frenzel_three-dimensional_2017}, desired asymmetry~\cite{sochava_chiral_1997,marques_role_2002}, or pronounced nonreciprocal properties~\cite{degiron_one-way_2014}. Such  arbitrary-illumination analysis, once developed, would be very beneficial for understanding and even design of bulk metamaterials and their two-dimensional counterparts (metasurfaces~\cite{yu_flat_2014,
glybovski_metasurfaces:_2016}) intended to interact with waves of arbitrary nature:  Acoustic,  electromagnetic, mechanical,  etc. 

In this paper, a universal platform for analysing properties of arbitrarily complex linear meta-atoms is proposed. 
Although we present the results for meta-atoms designed for interactions with  electromagnetic waves, the same approach can be used for the analysis of other types of wave interactions. We show that the response of a general linear meta-atom can be thought as a combination of responses of several basic modules with known, fundamental electromagnetic properties, immediately revealing the meta-atom properties for all possible excitations. Such a platform of transition from complex material units to basic elements (``materiatronics'') is in a sense analogous to electronics, where an arbitrarily complex linear circuit is represented as a network of basic elements: capacitors, inductors, resistors, and gyrators. There is also some analogy with recently introduced metatronics~\cite{engheta_circuits_2007} that relies on decomposition of complex electromagnetic systems into elementary scatterers with simple dispersion properties.

 The proposed modular analysis  is successfully  tested on several examples of dipolar meta-atoms with general electromagnetic properties, including bianisotropy and nonreciprocity. It is further demonstrated on the example of a split-ring resonator how the platform of materiatronics can be used for designing metasurfaces with  maximized (or  minimized) desired effects. For the convenience of the reader, the analysis approach developed in this paper is supported by a ready-to-use code and program files~\cite{suppl}. They provide automation in obtaining the  results for a meta-atom with arbitrary customer-defined geometry.

\section{Modular decomposition of polarization effects}
Let us first consider an arbitrary single  meta-atom located in free space so that its interaction with other meta-atoms can be neglected. Its electromagnetic response is assumed to be linear and dipolar, which is the case for  most inclusions of engineered metamaterials and metasurfaces.
No assumptions on its reciprocity or anisotropy properties are made, meaning that in the general case it can be bianisotropic and nonreciprocal. Thus, electromagnetic properties of the meta-atom are characterized by four polarizability dyadics (tensors): electric $\aee$, electromagnetic $\aem$, magnetoelectric $\ame$, and magnetic $\amm$, including in total 36 complex polarizability components. To apply the modular analysis of the meta-atom, one should first determine all these components. Several  techniques for polarizability extraction have been proposed during the last decade~\cite{arango_polarizability_2013,
asadchy_determining_2014,
yazdi_polarizability_2016,
liu_polarizability_2016,karamanos_full_2018} but none of them allows  calculation of all the polarizability components simultaneously. In this paper, the approach based on the scattered far field probing~\cite{asadchy_determining_2014} is extended~\cite{suppl} and utilized. It should be noted that even in the cases when the unknown meta-atom  cannot be described solely by dipolar moments, the modular analysis still can be introduced using alternative T-matrix formulation which takes into account higher-order multipoles~\cite{M_hlig_2011,fruhnert_computing_2017,
alaee_electromagnetic_2018}.

When the four polarizability dyadics of the meta-atom are determined, it is important to differentiate between the reciprocal   and possible nonreciprocal polarization properties. Using the Onsager-Casimir symmetry relations for bianisotropic meta-atoms~\cite{tretyakov_onsager-casimir_2002} $\aee (H_0)=\aee^T (-H_0)$,  $\amm (H_0)=\amm^T (-H_0)$,  $\aem (H_0)=-\ame^T (-H_0)$, one can represent each of $\aee$,  $\aem$,  $\ame$, and  $\amm$ dyadics  as a sum of the reciprocal  and nonreciprocal  parts (here, $T$ is the transpose operation, $H_0$ denotes all external nonreciprocal parameters such as bias magnetic field or angular velocity, and  $-H_0$ corresponds to the case when all these parameters switch signs). These reciprocal (subscript ``r'') and nonreciprocal (subscript ``n'')  parts read
\e\begin{array}{c}\displaystyle
\hspace{-1mm}\aeer=\frac{\aee+\aee^T}{2},            \hspace{9mm}\aeen=\frac{\aee-\aee^T}{2},   \\\displaystyle
\hspace{-1mm}\ammr=\frac{\amm+\amm^T}{2},             \hspace{3mm}\ammn=\frac{\amm-\amm^T}{2}, 
\end{array}\l{eq1}
\f\vspace{-4mm}
\e \begin{array}{l}\displaystyle
\hspace{-4mm}\aemr=\frac{\aem-\ame^T}{2},     
\hspace{6mm}\aemn=\frac{\aem+\ame^T}{2}. 
  \end{array}  \l{eq2} \f
The magnetoelectric dyadic $\ame$ is excluded from  Eq.~\r{eq1} since it is fully determined by    $\aem$. It should be noticed that $\aeer$ and $\ammr$ dyadics are symmetric, while  $\aeen$ and $\ammn$  are antisymmetric. However, both reciprocal and nonreciprocal electromagnetic dyadics have  symmetric and antisymmetric parts.

\begin{figure*}[ht!]
	\centering
	\subfigure[]{\includegraphics[width=0.24\linewidth]{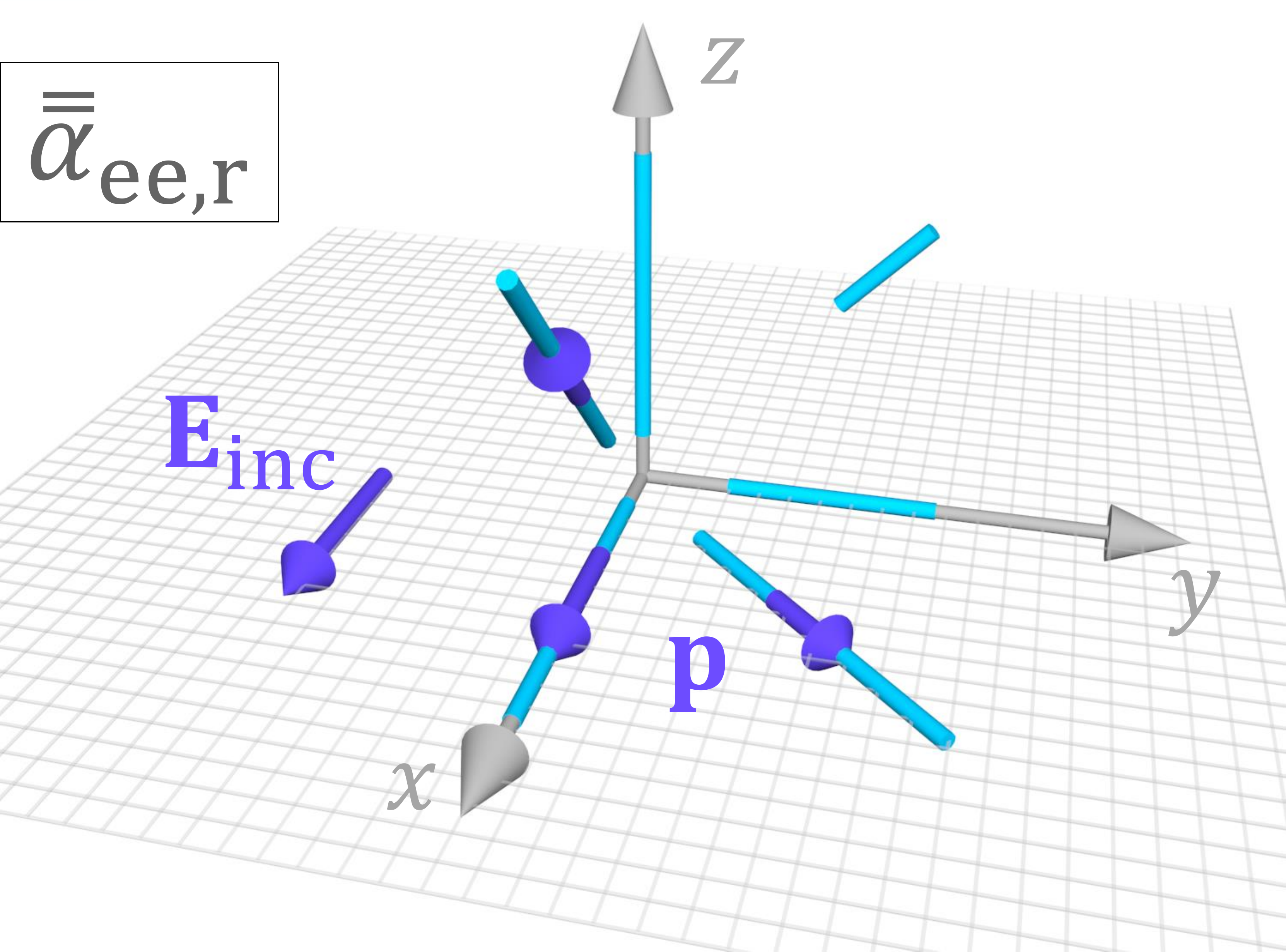} \label{fig1a}} 
	\subfigure[]{\includegraphics[width=0.24\linewidth]{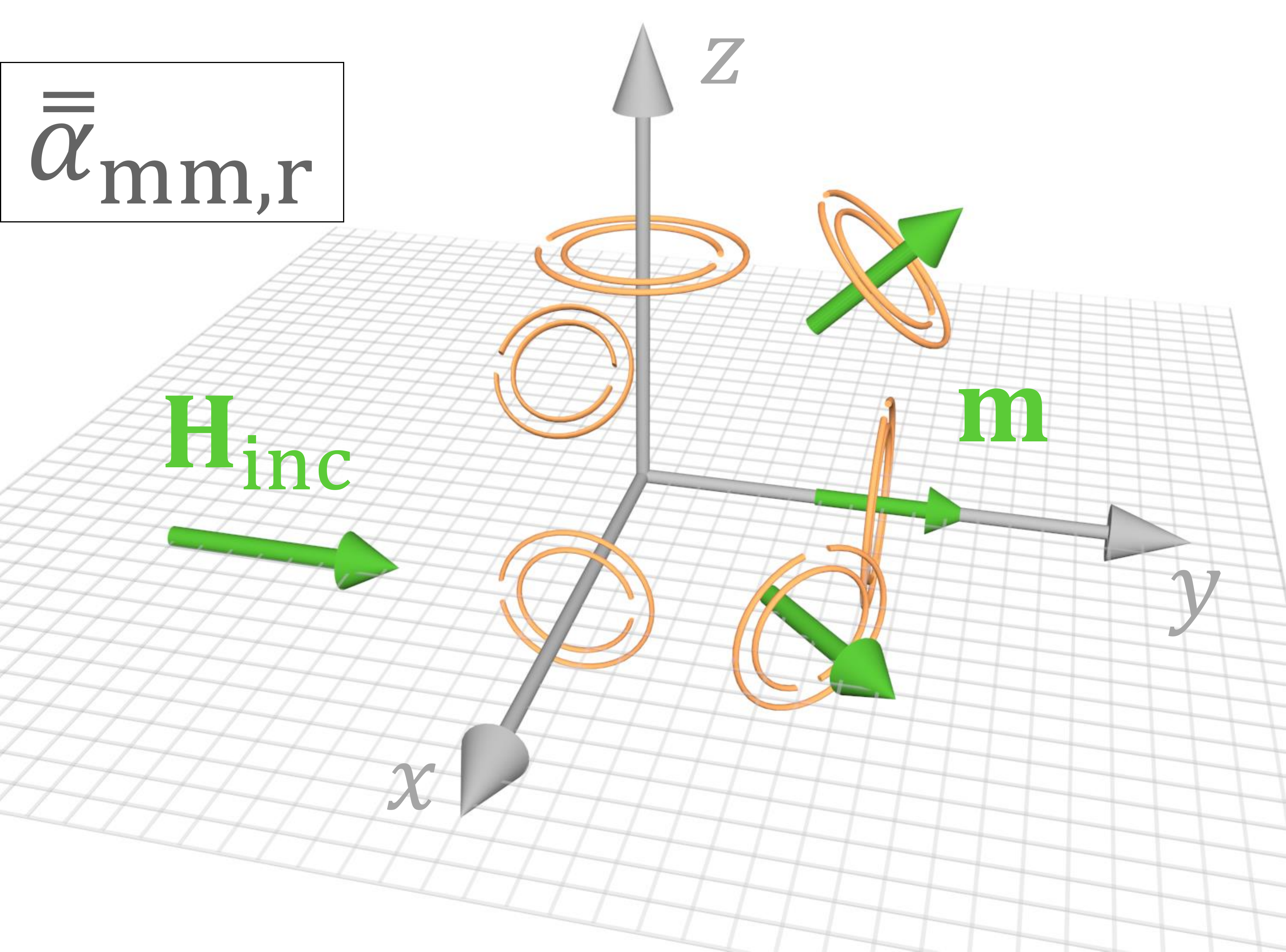}\label{fig1b}}
	\subfigure[]{\includegraphics[width=0.24\linewidth]{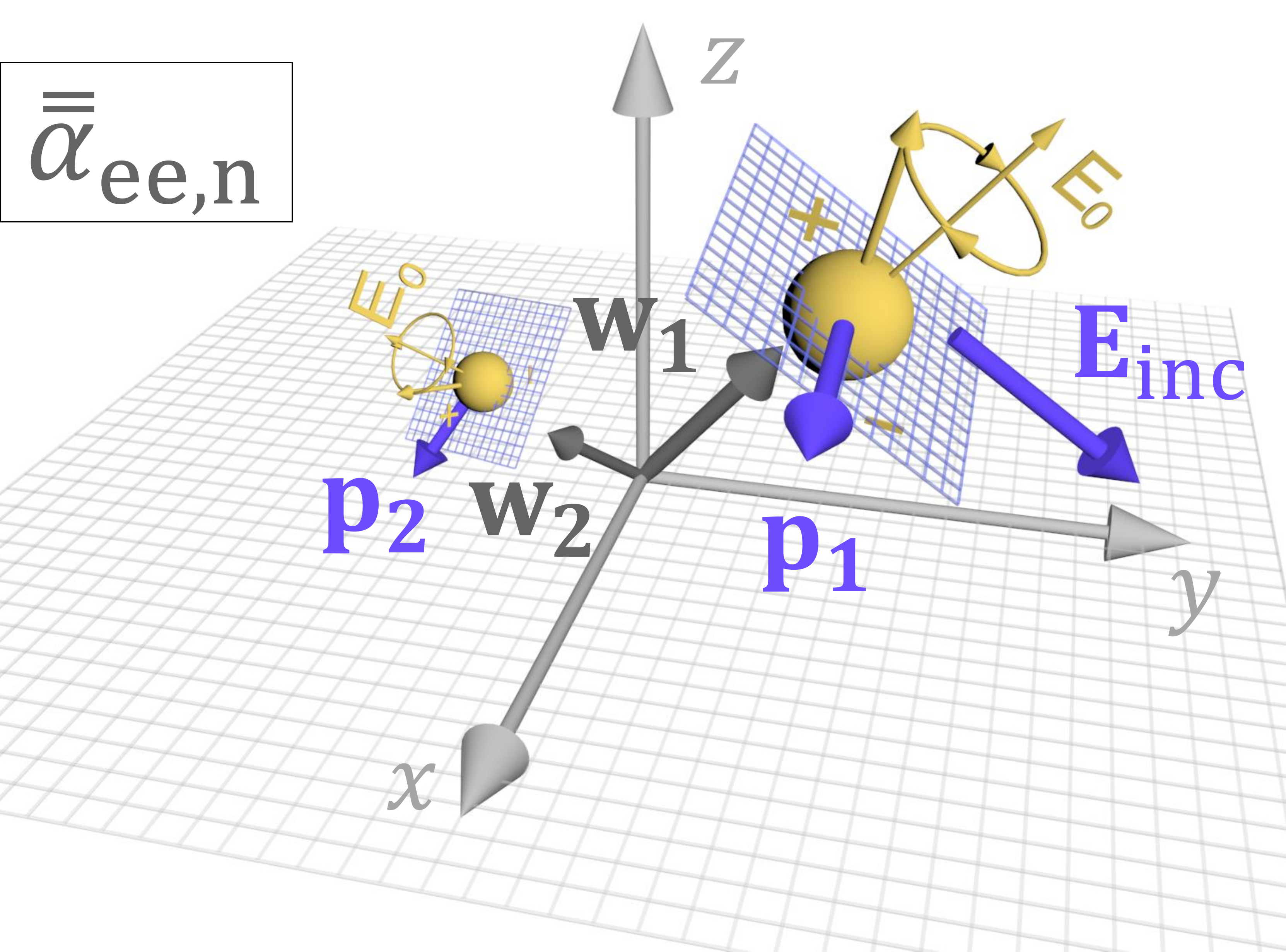} \label{fig1c}}
    \subfigure[]{\includegraphics[width=0.24\linewidth]{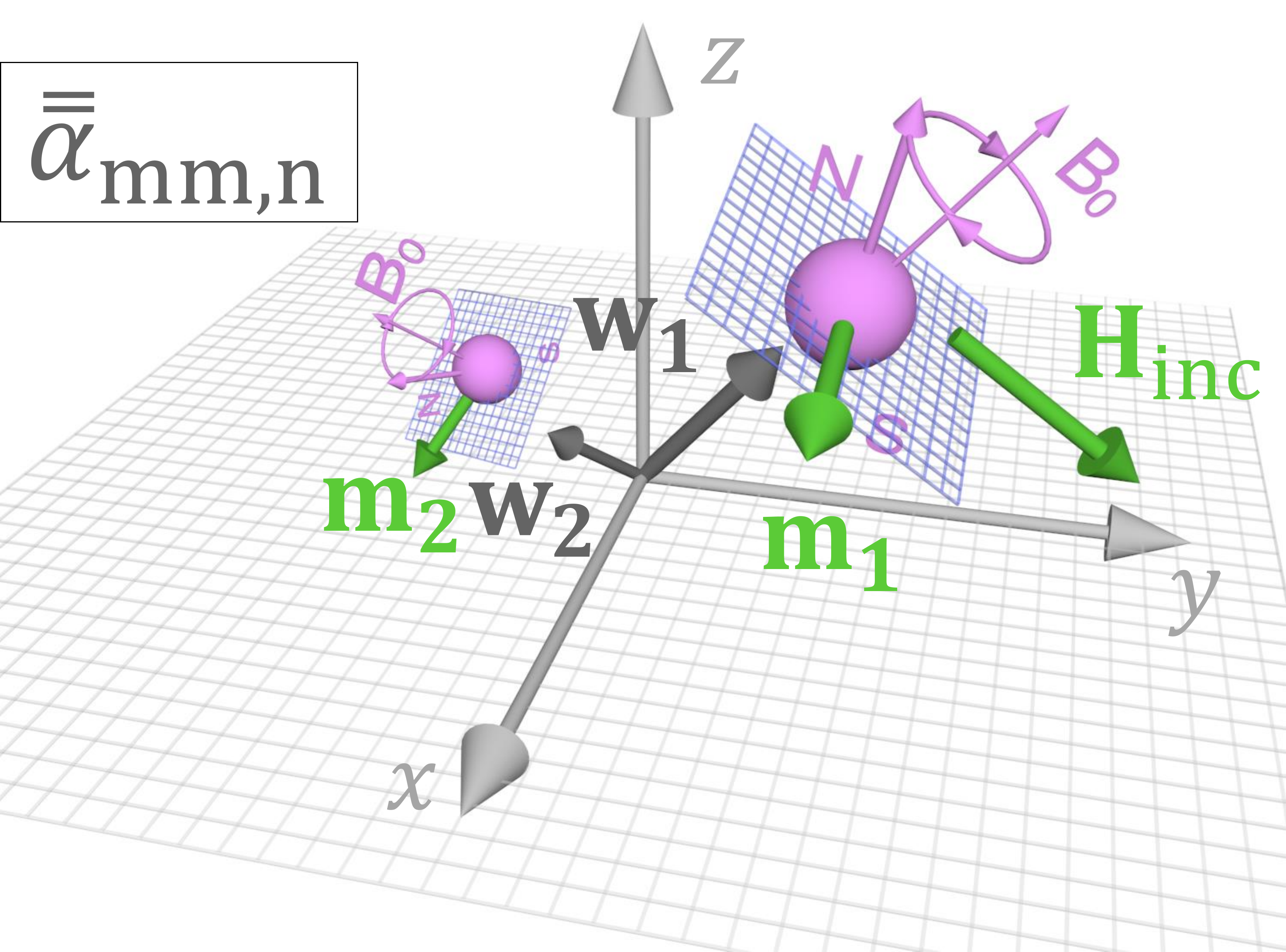} \label{fig1d}}\\
    	\subfigure[]
  {\includegraphics[width=0.24\linewidth]{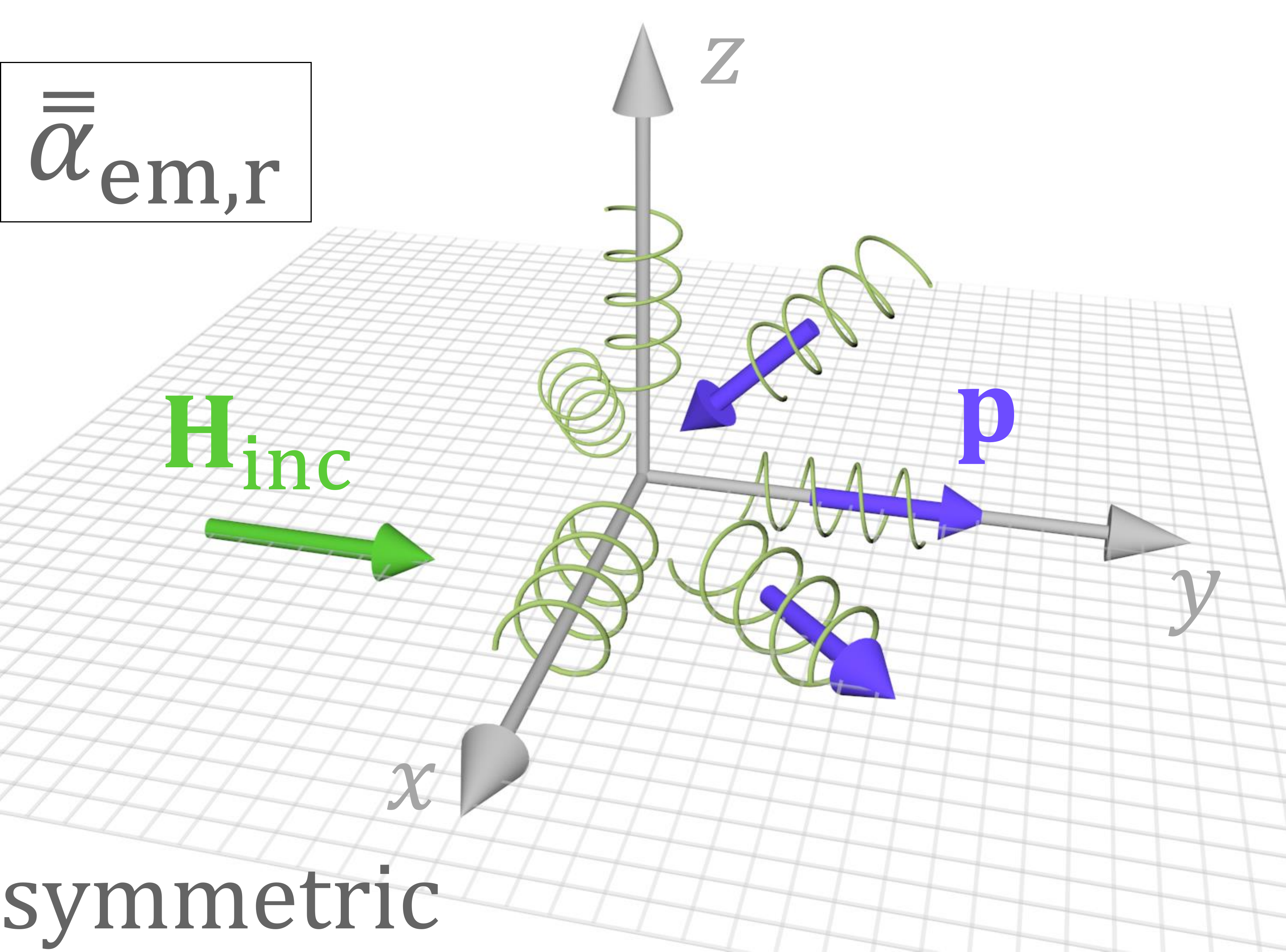} \label{fig1e}} 
	\subfigure[]{\includegraphics[width=0.24\linewidth]{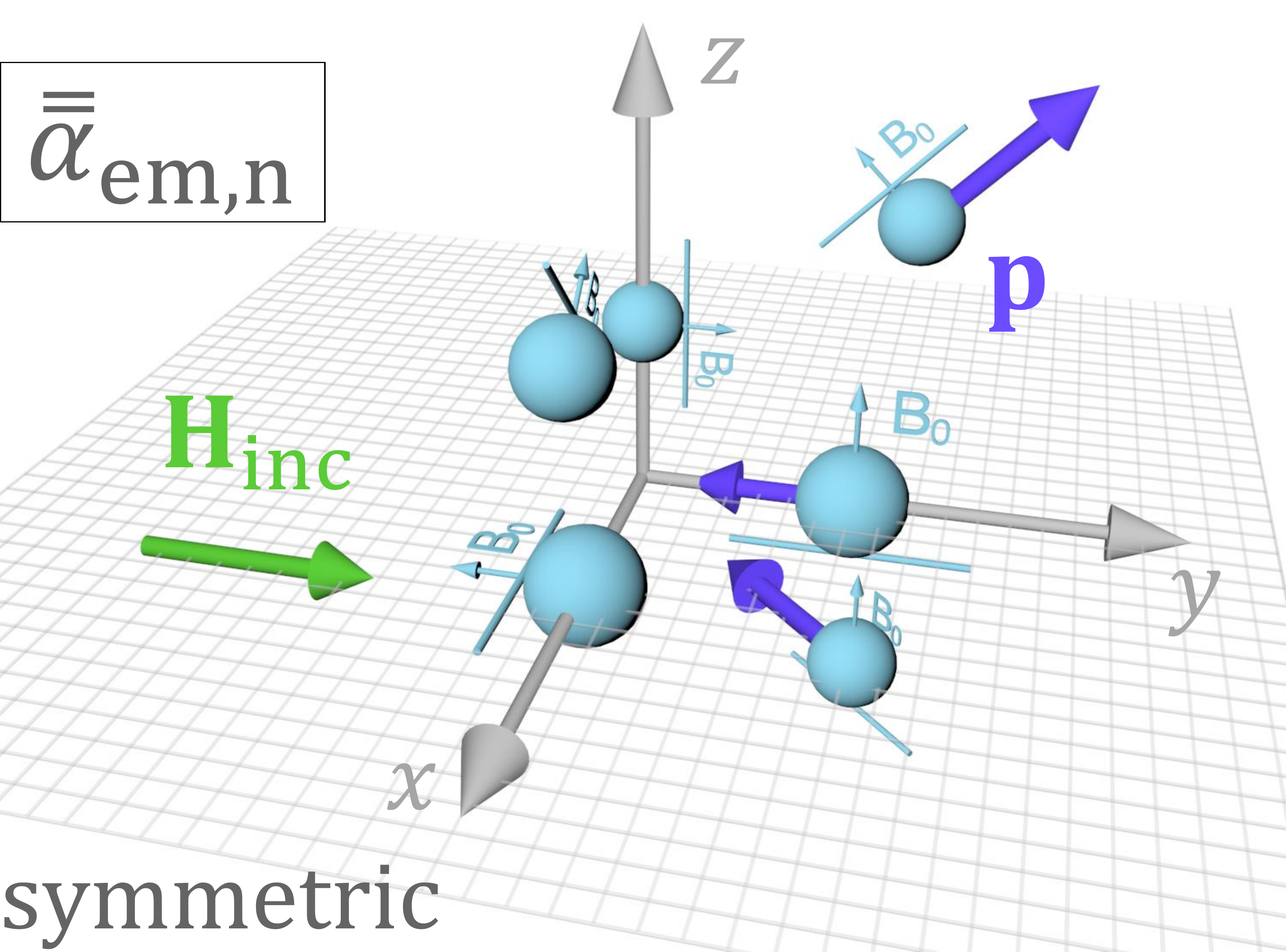}\label{fig1f}}
	\subfigure[]{\includegraphics[width=0.24\linewidth]{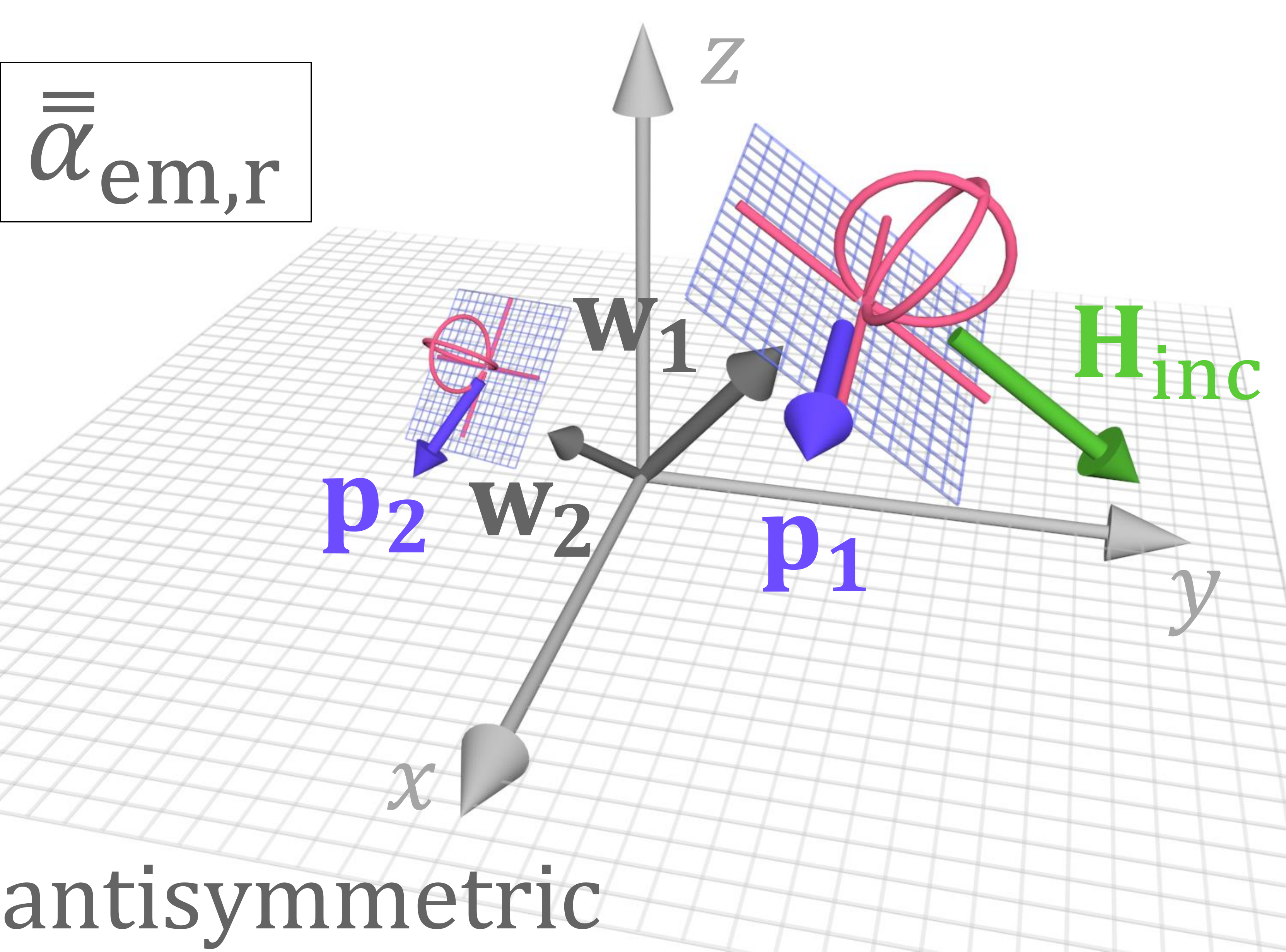} \label{fig1g}}
    \subfigure[]{\includegraphics[width=0.24\linewidth]{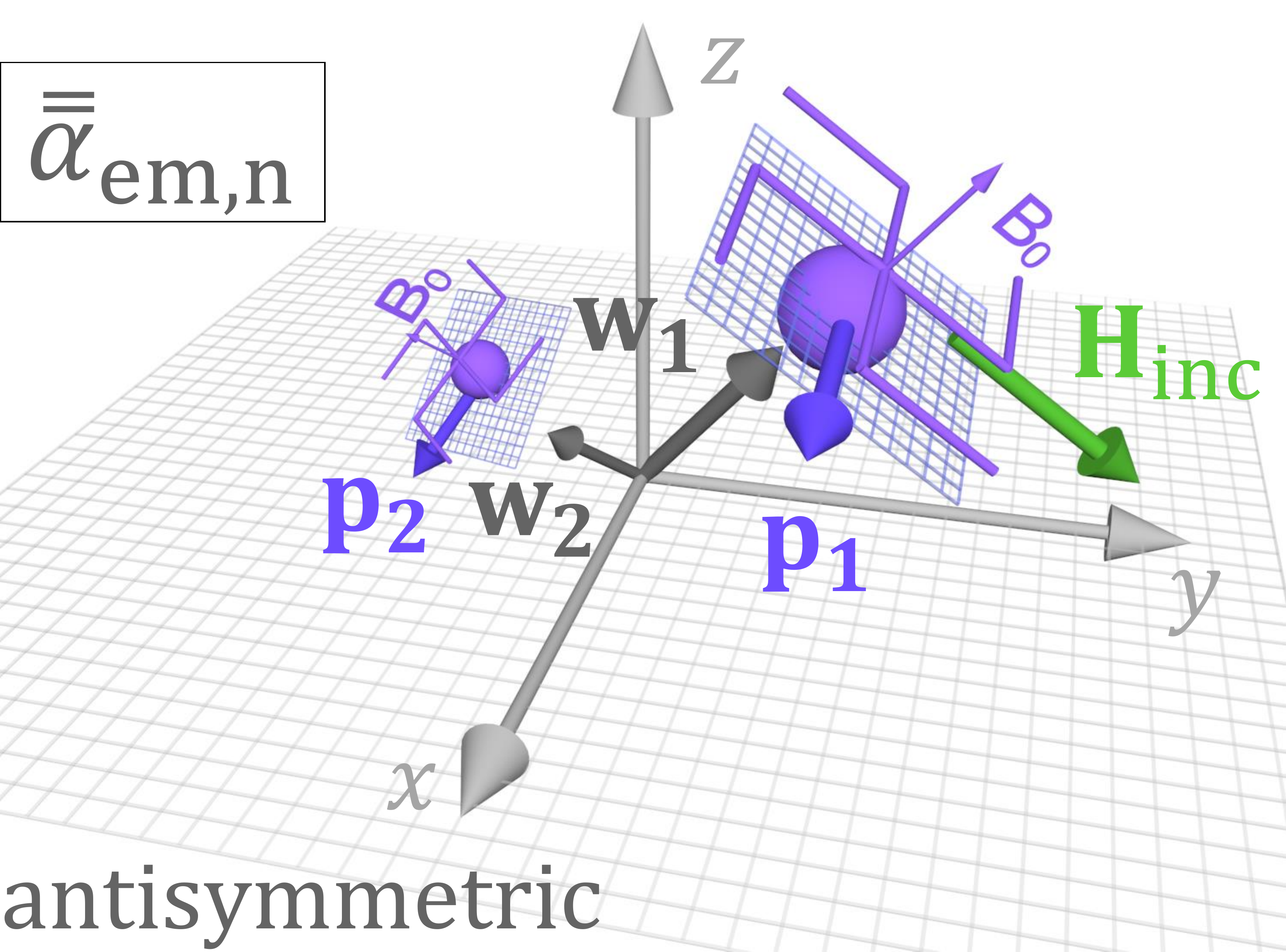} \label{fig1h}}
	\caption{ Conceptual illustration of modular decomposition of different polarizability dyadics for an arbitrary dipolar meta-atom. Incident electric and magnetic fields are denoted as $\textbf{E}_{\rm inc}$ and $\textbf{H}_{\rm inc}$, while $\textbf{p}$ and $\textbf{m}$ are the induced electric and magnetic dipole moments. Additional planes drawn at the antisymmetric modules are orthogonal to their symmetry axes. }
\end{figure*}

Any symmetric dyadic (e.g., $\aeer$)
 can be always represented as a linear combination of  a diagonal dyadic in the initial $xyz$-basis and another diagonal dyadic in another  basis given by complex basis vectors ${\bf u}_1$, ${\bf u}_2$, and ${\bf u}_3$~\cite{tai_vector_1997}: 
\e \aeer= \=I (\eexx+\eeyy+\eezz)/3 +\sum_{i} \lambda_i \_u_i\_u_i, 
\l{eq3}
\f
where $\=I=\_x \,\_x+\_y\,\_y+\_z\,\_z$ is the unit dyadic, $i=1,2,3$; $\lambda_i$ are complex coefficients such that $\sum_{i} \lambda_i =0$, and notation $\_u_i\_u_i$ denotes a dyadic product of   two vectors (the same as a dyad). The first diagonal dyadic in~\r{eq3} describes isotropic electric response. 
The second   dyadic  is diagonal with zero trace in the basis of complex unit vectors. For a clear geometrical description, it is convenient to rewrite it as a linear combination of six dyadic products of real vectors in the initial basis with complex magnitudes~$S_j$ ($j=1,2,..6$):
\begin{align}
\sum_{i} \lambda_i \_u_i\_u_i=&S_1\_x\,\_x +S_2\_y\,\_y+ S_3\_z\,\_z+ \frac{S_4}{2}(\_x+\_y)(\_x+\_y)   \nonumber \\
+ &\frac{S_5}{2}(\_x+\_z)(\_x+\_z)+ \frac{S_6}{2}(\_y+\_z)(\_y+\_z).
\l{eq4}
\end{align}
The complex amplitude coefficients $S_j$ can be  determined based on the extracted polarizability dyadics (see derivation in~\cite{suppl}).

Combining \r{eq3} and \r{eq4}, any symmetric dyadic in the general case is decomposed into six symmetric dyads (dyadic products of real vectors), all with different complex amplitudes. 
Thus, the response described by any reciprocal electric dyadic $\aeer$  is equivalent to the response of a set of six basic modules which possess only one simple polarization response: They develop electric dipole moment along a given direction when excited by electric fields along the same direction. For visualization purposes, we show  these modules as straight needles, as shown in Fig.~\ref{fig1a}, spatially separating them for clarity. 
Three  needles  are oriented along the initial basis vectors and the other three along the bisectors of the angles between these axes. It should be noted that such decomposition is not unique and in most cases, as will be seen below, several  of the modules might have negligible  weights, which reduces their total number. All the needles have different complex polarization amplitudes.  Hereafter, the amplitudes of the polarizability components are graphically shown as proportional to the \textit{linear} dimension of the module and not to its volume.  Now, the   polarization properties of a meta-atom with $\aeer$ given by~\r{eq3} can be readily determined from Fig.~\ref{fig1a}. We see that, for example, an incident $x$-polarized plane wave will induce polarization currents and dipole moments simultaneously in three modules, resulting in polarization of the meta-atom along the $x$, $y$, and $z$ axes (phases of polarization currents can be arbitrary).
In this way, we completely describe electromagnetic properties of any meta-atom only by the types of modules, their orientations and response strength of individual modules.



Similarly, symmetric dyadic $\ammr$   is decomposed into six dyads according to~\r{eq3} and \r{eq4} (naturally, in general,  with different complex amplitudes). Each decomposition term physically corresponds to a module which is reciprocal and polarizable only magnetically, only along one direction, and only by magnetic fields. For visualization purposes, we show the orientations and amplitudes of these modules by  double split-ring resonators~\cite{sauviac_double_2004,baena_equivalent-circuit_2005}, because this type of polarization response is dominant for this shape (obviously, the response of such resonators is more complicated, and the modular decomposition of an actual double split-ring resonator can be found in~\cite{suppl}). Thus, dyadic $\ammr$ can be modelled as the combination of six different idealistic double split-ring resonators, as illustrated in Fig.~\ref{fig1b}.

Any antisymmetric dyadics, such as $\aeen$, can be always represented as a cross product of a complex vector and the unit dyadic (complex vector is decomposed into two real vectors with real amplitude  $A_1$ and with imaginary amplitude $jA_2 $)~\cite{tai_vector_1997}:
\begin{align}
\aeen &  =\frac{1}{2} [(\eexy-\eeyx)(\_x\,\_y-\_y\,\_x)+
(\eexz-\eezx)(\_x\,\_z-\_z\,\_x)   \nonumber \\
+ &(\eeyz-\eezy)(\_y\,\_z-\_z\,\_y)]=
A_1 \textbf{w}_1 \times \=I +jA_2 \textbf{w}_2 \times \=I.
\l{eq5}
\end{align}
The unknown coefficients $A_1$ and $A_2$ and vectors $\textbf{w}_1$ and $\textbf{w}_2$ are calculated in~\cite{suppl}. The two dyadics in \r{eq5} describe  two uniaxial (with one symmetry axis) modules in which polarization occurs   in the direction orthogonal to the incident field and to  the  symmetry axis of the module. Physical interpretation of each such module in the case of nonreciprocal $\aeen$  dyadic is an electric dipole  precessing in a static external electric  field~$\textbf{E}_0$ (see Fig.~\ref{fig1c}). The induced electric dipole  is always perpendicular to the   electric field of the incident wave. Nonreciprocal magnetic dyadic $\ammn$, likewise, is modelled by two magnetic dipoles precessing in a static magnetic field~$\textbf{B}_0$ (see Fig.~\ref{fig1d}). 

\begin{figure*}[bt!]
	\centering
\includegraphics[width=0.98\linewidth]{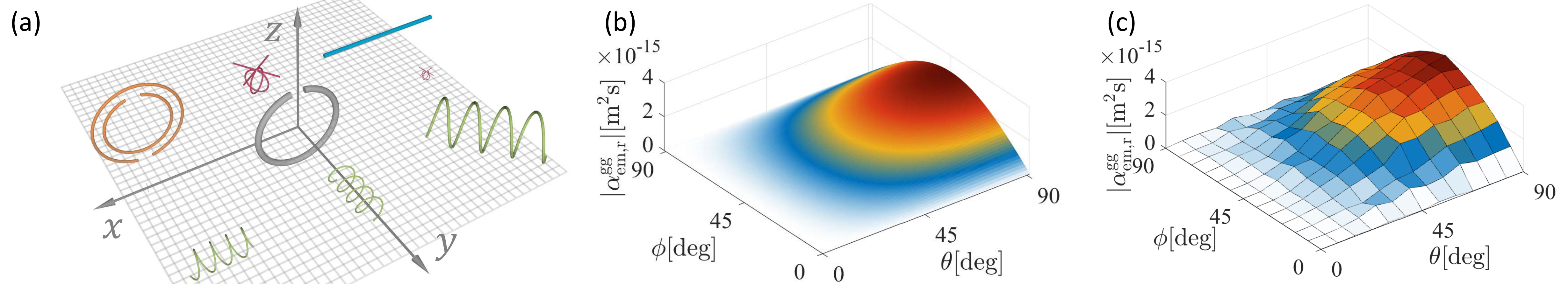} 
	\caption{ (a) Modular decomposition of a split-ring resonator at its fundamental resonance.   The linear dimensions of the modules are proportional to the amplitudes of the corresponding polarization effects. 
Surface plot illustrating the angular dependence of the amplitude of the chiral polarizability component $\alpha_{\rm em,r}^{gg}=\_g\cdot \aemr \cdot \_g$   of an SRR  meta-atom     with respect to the direction parallel to the unit vector  $\_g$.   Results obtained from the modular decomposition (b) and  results obtained using numerical full-wave simulations (c).
The maximum chirality is observed when the external electric field is oriented at $\theta=90^\circ$ and $\phi= 45^\circ$;  the SRR effectively behaves as a left-handed helix oriented at these angles.
	}
	\label{fig2}
\end{figure*}

Next, one should decompose the electromagnetic   dyadics $\aemr$ and $\aemn$ which have symmetric and antisymmetric parts and describe the bianisotropic response of the meta-atom. The  symmetric part of the former dyadic is equal to $(\aem+\aem^T-\ame - \ame^T)/4$ and according to \r{eq3} and \r{eq4} is modelled by six modules with uniaxial chiral bianisotropic response: For visual representation, thin multi-turn metal helices (see Fig.~\ref{fig1e}) are appropriate shapes to denote this fundamental polarizability effect. In general, we need two such chiral modules: Right-handed and left-handed. Incident magnetic field tangential to the helix axis generates electric polarization according to Faraday's law~\cite{Asadchy2018}.
Likewise, the symmetric part of  dyadic $\aemn$, equal to $(\aem+\aem^T+\ame +\ame^T)/4$, is represented by six uniaxial nonreciprocal  modules  (see Fig.~\ref{fig1f}). The effect of bianisotropic nonreciprocal coupling is strong in a meta-atom formed by a ferrite sphere covered by a metal wire (so-called Tellegen meta-atom)~\cite{tellegen_gyrator_1948,kamenetskii_technology_1996,
tretyakov_artificial_2003}, and we use an image of this meta-atom as a notation for the presence of this fundamental block in the decomposition. 
An incident alternating magnetic field creates cross-polarized magnetization   in the sphere (due to the off-diagonal permeability components of ferrite) which in turn results in induced electric polarization of the wire (parallel to the alternating magnetic field). 

The antisymmetric part of $\aemr$, expressed as $(\aem-\aem^T+\ame - \ame^T)/4$, according to \r{eq5} is modelled by two uniaxial bianisotropic asymmetric modules (one with real polarizability and another with imaginary polarizability), which we show as crossed omega-shaped wires~\cite{Saadoun_omega_1992}, see in Fig.~\ref{fig1g}. Incident magnetic field orthogonal to the module axis excites mutually orthogonal electric polarization (due to Faraday's law and   specific wire shape).  The last part in the modular decomposition is the antisymmetric  part of $\aemn$ that can be calculated as
$(\aem-\aem^T-\ame + \ame^T)/4$. Two nonreciprocal modules based on ferrite sphere with swastika-shaped metal wires~\cite{tretyakov_nonreciprocal_1998} (shown in  Fig.~\ref{fig1h}) can represent the required polarization properties. 

\section{Modular decomposition for a split-ring resonator}
Next, as all the polarizability dyadics $\aee$, $\aem$, $\ame$, and $\amm$ are decomposed into basic modules,  the universality of the materiatronics platform can be demonstrated via modular analysis of  realistic meta-atoms. The analysis for each of them was performed in three automated steps: Extraction of the full polarizability dyadics of the unknown meta-atom~\cite{suppl} (performed using a full-wave simulator~\cite{hfss}), decomposition of the extracted polarizability dyadics (using computational software~\cite{matlab}), and  three-dimensional visualization of the obtained modules (using the full-wave simulator). All the supporting files for the ready-to-use automated  analysis of an unknown meta-atom  can be found in~\cite{suppl}.

First, the modular analysis is applied to a split-ring resonator (SRR), as an example.   The geometry and orientation of the SRR (shown in grey) in the initial basis can be seen in Fig.~\ref{fig2}(a). The radius of the  ring is 2.7~mm, the radius of the copper wire is 0.2~mm, and the gap is 0.4~mm. The decomposition was performed at the main resonance frequency of the SRR at 7.9~GHz. The power of the modular analysis is evident from Fig.~\ref{fig2}(a) where the electromagnetic response of the meta-atom can be immediately inspected for arbitrary illuminations. The strong electric polarization response along the $x$-axis (straight wire) and magnetic polarization along the $y$-axis (magnetic module shown as a double SRR) are expected. However, additionally, the SRR exhibits perceptible omega bianisotropic response when illuminated along the $z$-axis~\cite{sochava_chiral_1997,
marques_role_2002} with specific polarization of the incident wave~\cite{note}. Moreover, the decomposition includes two left-handed and one right-handed helices, indicating strong chiral properties of the SRR. 
Although the total chirality is zero (left- and right-handed helices precisely compensate each other; vanishing three-dimensional chirality), for specific illumination directions  one can excite predominantly left- or  right-handed helices, achieving non-zero chiral effect with a planar structure (the SRR can be \textit{infinitesimally} thin supporting only electric current). The planar chirality was observed earlier in various asymmetric structures~\cite{bunn_chemical_1961,
williams_opticalrotatory_1969,
sochava_chiral_1997,
papakostas_optical_2003,
plum_metamaterials:_2009}, however, this effect remained rather weak~\cite{plum_metamaterials:_2009,
singh_highly_2010}. Using the modular decomposition, one can determine the chirality  strength exhibited by the SRR for various orientations of the incident electric field. 
As it is theoretically shown in~\cite{suppl}, the maximum chirality is achieved when the electric field is along 
unit vector  $\_g= \sin \theta \cos \phi \_x+ \sin \theta \sin \phi \_y + \cos \theta \_z$,  where $\theta=90^\circ$ and $\phi=\pm 45^\circ$ (the double sign is due to the mirror symmetry of the SRR with respect to the $yz$-plane). 
Figure \ref{fig2}(b) depicts 
 theoretically calculated axial chiral polarizability of the split-ring resonator for different orientations of the incident electric
field. Indeed, the maximum of chirality is observed  at $\theta=90^\circ$ and $\phi= 45^\circ$. 
To verify this theoretical result, polarizability $\aemr$ was calculated using the above mentioned extraction   technique, described separately for each  orientation of the incident electric field in the range of $\theta=[0^\circ;90^\circ]$ and  $\phi=[0^\circ;90^\circ]$ with the step of $7.5^\circ$.
The results are depicted in Fig.~\ref{fig2}(c) and are in close agreement with the theoretical ones in Fig.~\ref{fig2}(b). The deviation in the peak amplitude value between these two figures (around 30\%) can be explained by errors (which were of the order of 26\% for this meta-atom) in the polarizability extraction.

As can be seen from Fig.~\ref{fig2}(a), the SRR behaves differently for opposite illumination directions: As a left-handed helix when illuminated along the ($\_x- \_y$) direction (the right-handed helix in the decomposition in Fig.~\ref{fig2}(a) is not excited) and as a right-handed helix when illuminated along the ($-\_x - \_y$) direction. 

\section{Design of negligibly thin metasurface with extreme circular dichroism }
\begin{figure}[tbh]
	\centering
\includegraphics[width=0.88\linewidth]{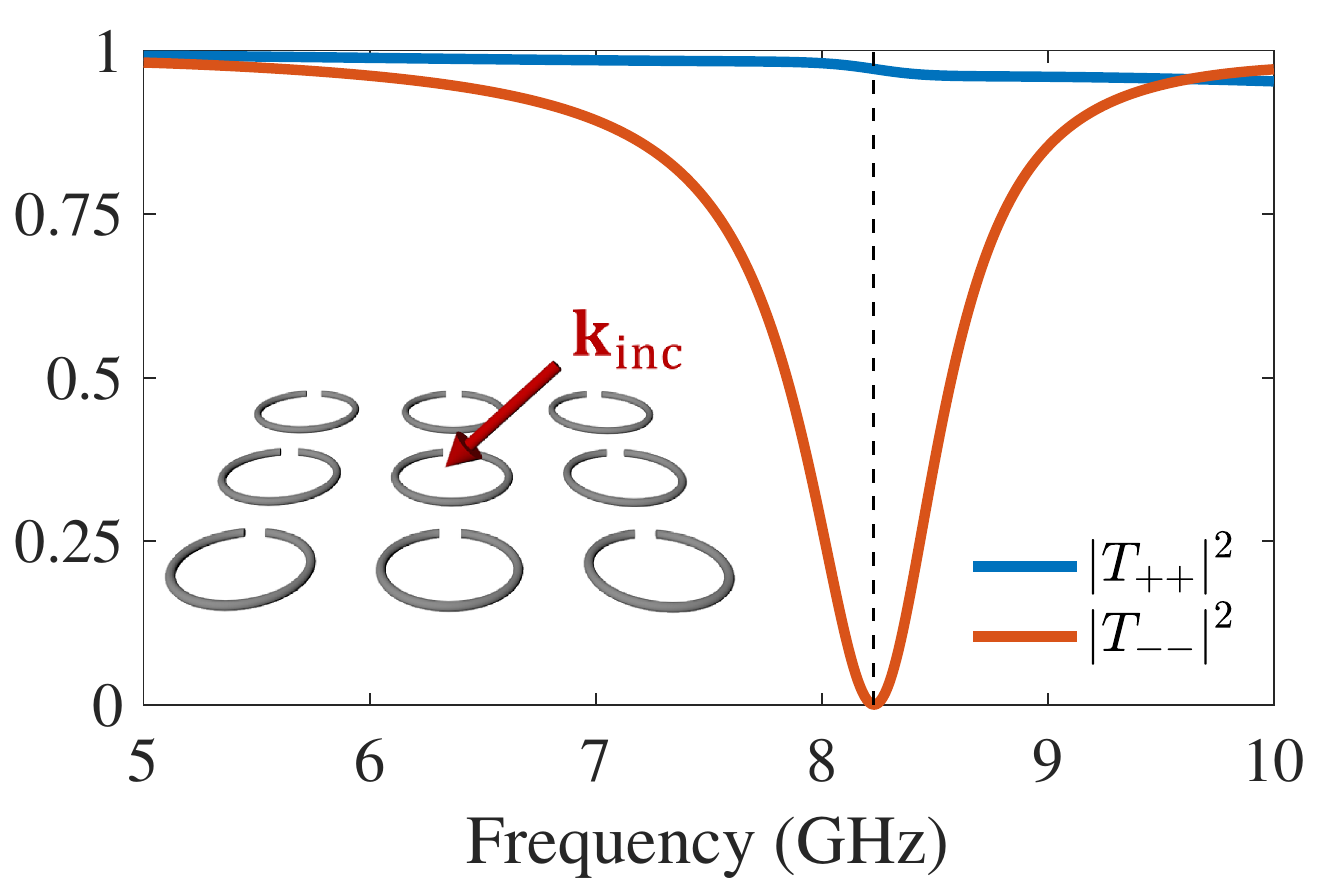} 
	\caption{  Transmittance through a metasurface composed of SRRs for right (``$++$'') and left (``$--$'') circularly polarized incident waves. The inset depicts the geometry of the metasurface and wave vector $\textbf{k}_{\rm inc}$ of the incidence.
	}
	\label{fig3}
\end{figure}

\begin{figure}[tbh]
	\centering
\includegraphics[width=0.78\linewidth]{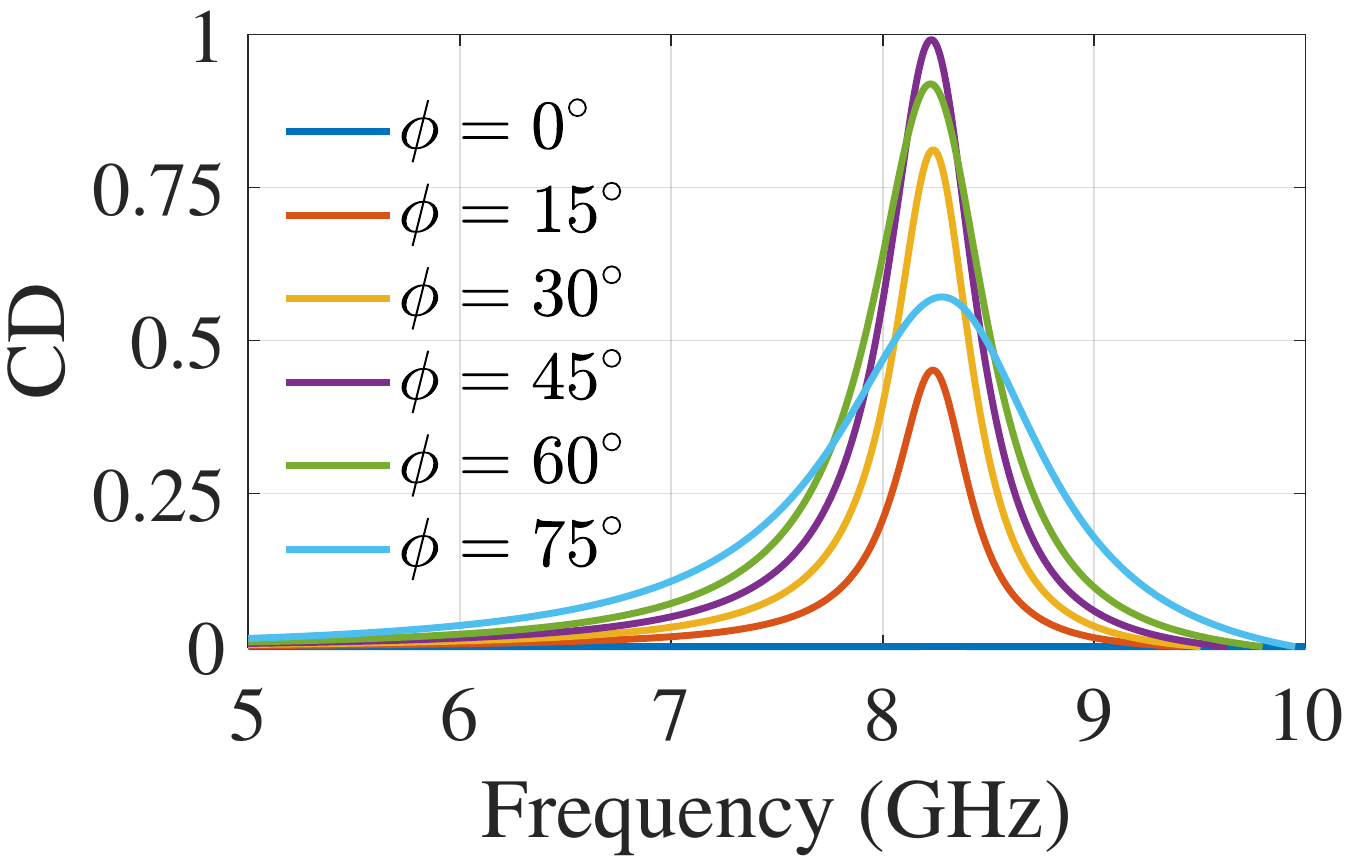} 
	\caption{Circular dichroism of the metasurface versus frequency for different  illumination angles $\phi$. }
	\label{fig4}
\end{figure}

The applicability of the modular analysis can be further demonstrated by designing a periodic metasurface of SRRs with the  dimensions listed in the previous Section. The metasurface has a subwavelength square unit cell with the $10$~mm side. The  SRRs are \textit{parallel} to the metasurface plane. In contrast to conventional chiral metasurfaces relying on finite thickness and three-dimensional topology with broken space-reversion symmetry, the designed metasurface exhibits strong circular dichroism (and other chiral effects) with pure two-dimensional geometry and negligible thickness.
The metasurface was illuminated by circularly polarized plane waves at an angle $\phi=45^\circ$ from the normal [${\_k}_{\rm inc} \uparrow\uparrow (\_x - \_y)$ in the initial basis], as shown in the inset of Fig.~\ref{fig3}(c). For this illumination, the right-handed helix shown in Fig.~\ref{fig2}(a) is not excited, while both left-handed helical modules are activated. Thus, the SRR behaves for this illumination as a left-handed chiral meta-atom.
The plot in the figure shows frequency dispersions of the transmittance  for right (``$++$'') and left (``$--$'') circular polarization of incident waves from full-wave simulations. At the  resonance frequency of 8.23~GHz, the metasurface nearly fully transmits incident right circular polarization  and completely reflects waves of the opposite handedness (transforming their polarization). Cross-polarization transmittance ($|T_{+-}|^2$ and $|T_{-+}|^2$) of the metasurface is indistinguishable from zero within the entire studied frequency range. Thus, circular dichroism defined as ${\rm CD}=(|T_{++}|^2+|T_{-+}|^2-|T_{--}|^2-|T_{+-}|^2)/ (
|T_{++}|^2+|T_{-+}|^2+|T_{--}|^2+|T_{+-}|^2)$ is equal to~1, which is an exceptional result taking into account negligible thickness of the metasurface. Figure~\ref{fig4} shows CD of the designed metasurface   for different illumination angles. The maximum value of CD is achieved for $\phi=45^\circ$, while for normal incidence ($\phi=0^\circ$) it is zero.
 
Such high CD significantly exceeds the values reported in the literature for planar structures to date~\cite{plum_metamaterials:_2009,
singh_highly_2010}. Furthermore, the effect of planar chirality achieved with the designed metasurface should be distinguished from the anisotropy effect observed in planar dielectric metasurfaces~\cite{wu_spectrally_2014}. In the latter case, a metasurface exhibits   symmetric response in terms of co-polarized transmission $|T_{++}|=|T_{--}|$ and asymmetry for cross-polarized transmission  $|T_{+-}|\neq|T_{-+}|$.

\section{Modular decomposition for a nonreciprocal virtually moving meta-atom}
\begin{figure}[tb!]
	\centering
\includegraphics[width=0.98\linewidth]{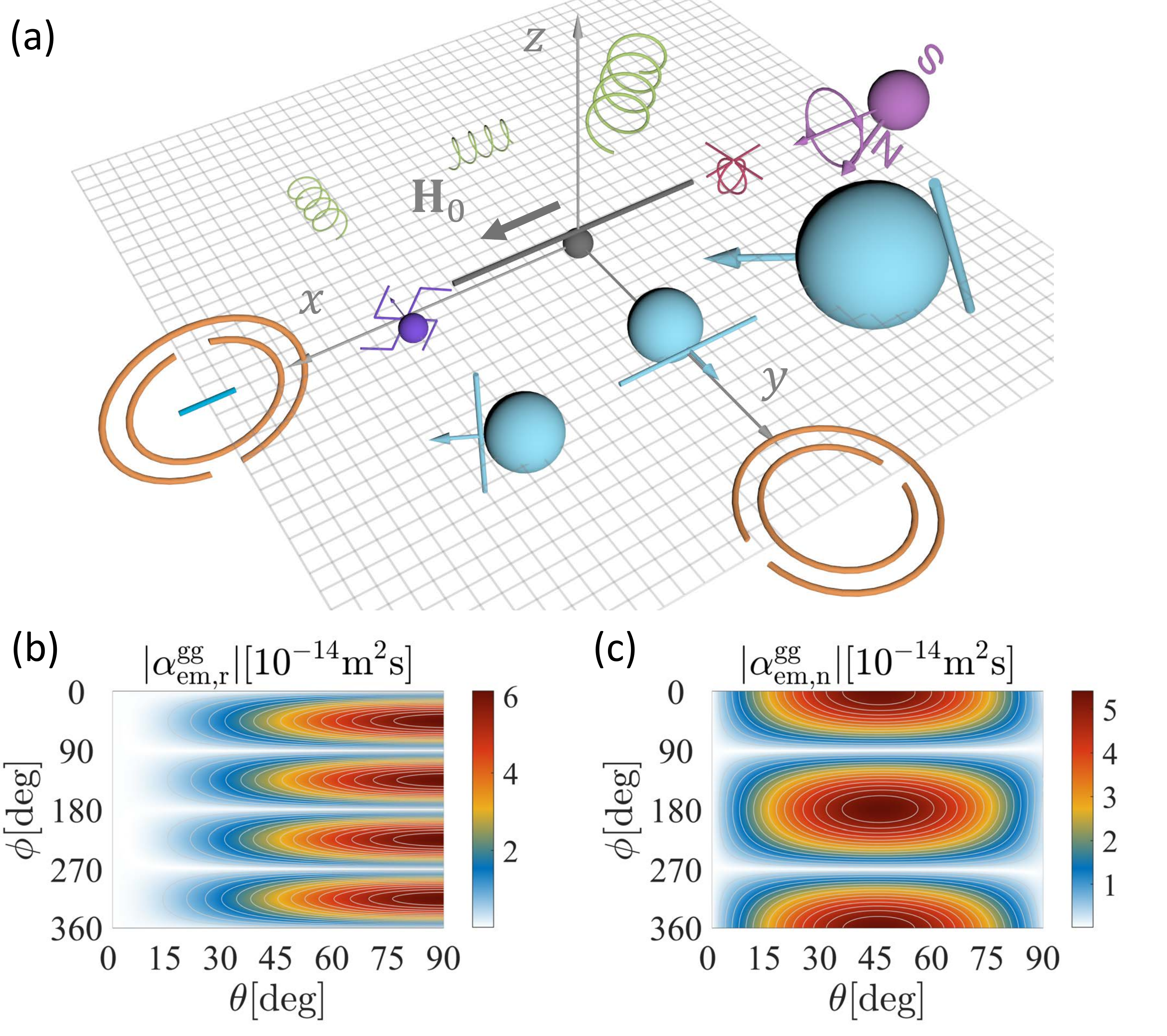} 
	\caption{(a) Modular decomposition of a   nonreciprocal meta-atom  (shown in grey)  at its magnetic resonance frequency.     Surface plots depicting  axial (b) chiral and (c) Tellegen polarizabilities  for different orientations of the incident electric field in the spherical coordinate system. The maximum chirality is observed when the electric field is oriented at $\theta=90^\circ$ and $\phi= (45^\circ+90^\circ N)$, and the maximum Tellegen properties for $\theta=45^\circ$ and $\phi= 180^\circ N$. }
	\label{fig5}
\end{figure}

In this section, modular decomposition is applied to a nonreciprocal meta-atom which can be used as a constituent of metasurface-based isolators and phase shifters. The geometry of the meta-atom was adopted from~\cite{degiron_one-way_2014} and is shown in Fig.~\ref{fig5}(a).
The length of the copper wire is 30~mm, and the wire radius is 0.05~mm. The ferrite sphere of 1.65~mm radius is made of yttrium iron garnet with the relative permittivity of~15, the dielectric loss tangent $10^{-4}$, the saturation magnetization  $1780$~G, and the full resonance linewidth $0.2$~Oe. The meta-atom is biased by an external $+z$-oriented permanent   magnetic field of $9626$~A/m.   The magnetic resonance frequency at which the modular analysis was performed is 1.975~GHz. The orientation  of the magnetic bias field is along the $+x$-direction. 

As it was theoretically demonstrated in~\cite[Eqs. (11--12)]{radi_one_way_2014}, \cite{
taravati_nonreciprocal_2017}, \textit{passive} metasurfaces operating as highly efficient nonreciprocal phase shifters or isolators at normal incidence must consist of meta-atoms  which possess predominantly artificial ``moving'' bianisotropic coupling (i.e.,  the entire dyadic $\overline{\overline{\alpha}}_{\rm em}$ must be represented solely by the nonreciprocal antisymmetric part  $\overline{\overline{\alpha}}_{\rm em, n}$). Such result has a simple conceptual explanation. A perfect nonreciprocal isolator (lossy device) or   phase shifter (lossless device) must be fully transparent from one side, meaning that they do not modify neither the amplitude nor the phase of incident waves. Such functionality may seem impossible because it implies the absence of any wave scattering from the metasurface, while currents in the meta-atoms are non-zero. However,  in an artificial ``moving'' metasurface,  currents induced due to   bianisotropic effects and due to electric and magnetic polarizations of the meta-atoms  cancel scattering from each other and form a nonscattering system. Conceptually, such metasurface resembles a  thin layer with homogeneous and isotropic electric and magnetic properties   moving outwards from the source of incident waves with the  speed of light. This way we emulate the situation when the incident waves simply cannot reach the layer, yielding zero scattering. When illuminated from the opposite direction, the layer will strongly scatter waves.

Modular decomposition shown in Fig.~\ref{fig5}(a) allows us to immediately understand if the  nonreciprocal meta-atom under analysis in fact has  necessary  properties. At   first glance, it seems that the meta-atom owns polarization properties  of all possible kinds and, therefore, is not suitable for implementation in metasurface-based isolators. 
However, if one assumes 
 the illumination   of the meta-atom   along the $+y$-direction with the polarization of the incident fields given by  $\textbf{E}_{\rm inc}=-\textbf{x}E_{\rm inc}$ and $\textbf{H}_{\rm inc}=\textbf{z}H_{\rm inc}$ (same assumption was made in~\cite{degiron_one-way_2014}), the conclusion becomes opposite. From the modular decomposition, one can make the following observations. Firstly, the artificial ``moving'' module is excited at most since the illumination is along its axis. Secondly,   the twisted omega module is not excited since the illumination is perpendicular to its axis. Thirdly,   taking into account the orientation of $\textbf{E}_{\rm inc}$ ($\theta=90^\circ$ and $\phi= 180^\circ$), the chiral and Tellegen bianisotropic effects are minimized. The latter observation can be also confirmed by the plots in Figs.~\ref{fig5}(b) and (c) depicting   axial chiral and Tellegen   polarizability components for different orientations of the incident electric field. The maximum of chiral response is observed for $\theta=90^\circ$ and $\phi= (45^\circ+90^\circ N)$ ($N$ is an integer), while the maximum of Tellegen properties is achieved for $\theta=45^\circ$ and $\phi= 180^\circ N$.

Thus, modular decomposition  of meta-atoms is a useful method to determine their   polarization properties for all illumination directions. 
Three  other examples of  meta-atom decomposition, including double-turn helix, double split-ring resonator, and nonreciprocal swastika-shaped inclusion, can be found in Supplementary Materials~\cite{suppl}.

\section{Conclusions}
The above examples  clearly demonstrate the usefulness and potential of the materiatronics platform. Using relatively simple vector algebra manipulations, one can analyze the polarization properties of an arbitrary meta-atom for all possible configurations of incident plane waves, identify \textit{all} possible (under the assumption  of linear dipolar response) scattering effects in this meta-atom, and find optimal excitations for maximizing or minimizing particular effects.
 The proposed analysis can be exploited not only for analysis of existing meta-atoms and artificial composites based on them, but also as a starting point for meta-atom synthesis. Indeed, by tuning geometrical parameters of a meta-atom and analysing the resulting modifications of the conceptual modules in its decomposition, one can gain an insight about how to engineer necessary polarization effects. 
Although the modular analysis reported in this paper was introduced based on the polarizability description, it is possible to extend it analogously to susceptibility characterization, which is beneficial for metasurfaces incorporating metal-backed dielectric layers.




\section*{Acknowledgment}

This work was supported by the Finnish Foundation for Technology Promotion  and Academy of Finland (project 287894).

\bibliography{bibliography}
\bibliographystyle{unsrt}
\end{document}